# The impact of sensory characteristics on the willingness to pay for honey

Zaripova J.[1], Chuprianova K.[1†], Polyakova I.[1,2†], Semenova D.[1], Kulikova S.[1*]

1. Center for Cognitive Neuroscience, HSE University, Perm, Russia.

2. The International Laboratory of Intangible-driven Economy, HSE University, Perm, Russia.

† equal contribution

* for correspondence: SPKulikova@hse.ru
  37A, blvd. Gagarina, Perm, Russia, 614060
  Center for Cognitive Neuroscience, HSE University

**Abstract.** Honey consumption in Russia has been actively growing in recent years due to the increasing interest in healthy and environment-friendly food products. However, it remains an open question which characteristics of honey are the most significant for consumers and, more importantly, from an economic point of view, for which of them consumers are willing to pay. The purpose of this study was to investigate the role of sensory characteristics in assessing consumers' willingness to pay for honey and to determine which properties and characteristics "natural" honey should have to encourage repeated purchases by target consumers. The study involved a behavioral experiment that included a pre-test questionnaire, blind tasting of honey samples, an in-room test to assess perceived quality, and a closed auction using the Becker-DeGroote-Marschak method. As the result, it was revealed that the correspondence of the expected sensations to the actual taste, taste intensity, duration of the aftertaste and the sensations of tickling in the throat had a positive effect on both the perceived quality of the product and the willingness to pay for it, while perception of off-flavors or added sugar had a negative impact. Using factor analysis, we have combined 21 sensory characteristics of honey into eight components that were sufficient to obtain the flavor portrait of honey by Russian consumers.

**Keywords:** honey, perception, sensory characteristics, willingness-to-pay, factor analysis

# 1. Introduction

Honey and other apiculture products are the fairly popular categories, regularly found in the food basket of modern Russian consumers. Researchers note a growing trend in honey consumption, caused by the increasing interest in healthy food products of organic origin. According to the Russian Agricultural Bank (2023), a typical Russian citizen consumes ~0.5kg of honey per year .

Honey is positioned as a healthy food product due to its composition. It is commonly used as a substitute for refined sugar in hot and cold beverages, as a traditional ingredient in cooking cakes, desserts, sauces, sandwiches, and for accompanying cheese (Kowalczuk et al., 2023). Antioxidants, phenols, and flavonoids, that are present in honey, may have favorable effects on mental health, contribute to memory and effective stress management (Zamri et al., 2023). The motives for consuming honey are diverse. The product is widely discussed as an adjuvant folk remedy to treat upper respiratory tract diseases and reduce body temperature, heal wounds, support the immune system, lower cholesterol levels and even to prevent some forms of cancer (Kumar & Bhowmik, 2010). Honey is also used for cosmetic purposes to moisturize, purify and improve skin elasticity, to relieve irritation and slow down the aging process. Modern consumers also actively mention the nutritional properties of honey (Grontkowska & Grzyb, 2019).

In Russia, honey production increased annually by approximately 1% from 2017 to 2022 (Kostenko, 2022). This growth rate was determined by the significant increase in the number of households and farms over the same five year period. Evolving competition among honey entrepreneurs stimulates companies to take into account individual product attributes that represent a sustainable market advantage in a rapidly changing external environment. Among the key factors potentially influencing consumers' willingness to pay (WTP) for apicultural products, experts emphasize honey's eco-friendliness and naturalness (Vapa-Tankosic et al., 2020). Trends in maintaining healthy lifestyles have further spotlighted these attributes. Legislative innovations aimed at preserving bee families and facilitating conscientious beekeeping practices are also driving growth in the organic honey production. In the present work, we have adressed the following interrelated questions:

- Does the concept of "naturalness" increase consumer willingness to pay for a product?
- Which honey characteristics positively influence perceived product quality?
- Which of the following qualities encourage repeated purchases by target consumers: sweet or bitter, liquid or crystallized, strong or weak aftertaste?

Within this study, we were primarily focused on the consumption of honey as a widespread food product. Before describing the methodology of the behavioral part of this work, it is important to consider some aspects of the consumers' perception of honey, specifically a set of factors influencing its perception and the role of sensory characteristics in assessing consumers' willingness to pay for it.

## 2. Theoretical background

### 2.1. Factors influencing the honey purchase decision

Honey is a "complex" food product and the decision to purchase it is influenced by a wide range of factors (Sparacino et al., 2022). Previous research has identified the three key variables influencing consumers' willingness to pay, namely: distribution channel, packaging, and price. Each of these factors is described below.

*Distribution channel.* The majority of honey consumers prefer to buy it directly from farmers. According to Kowalczuk et al. (2017), 32% of the population look for a product in open-air markets, and 27% in apiaries. These are representatives of the older generation aged 45-74 years, who trust only beekeepers they know personally. On the contrary, population under 30 years of age, characterized by a relatively low awareness of methods for choosing a high-quality product, buy honey in supermarkets. According to Rozdolskaya et al. (2015), only 4.54% of Russian consumers purchase honey in retail chains. As the researchers observed, most buyers prefer to get honey from a beekeeper or, in the absence of a reliable seller, at specialized fairs. Interestingly, the sales channel influences perception of the honey sensory characteristics.

Consumers perceive honey bought from a beekeeper as "tastier and more flavorful" than honey in a store (Roman et al., 2013). However, the impact of the distribution channel may differ across regions and countries. For example, according to Hunter et al. (2021), 70% of Australian consumer trust the quality of a product bought in the supermarket. Given the trend towards healthy lifestyles among the Russian population and the inevitable rejuvenation of the product's target audience, domestic producers should concern about designing the product for successful launch on the supermarket shelves.

*Packaging.* The importance of packaging for choosing honey is not as obvious as it may seem from the first sight. On the one hand, packaging design influences the perceived quality of the product and on average, people are willing to pay more for products with an attractive visual design (Setiowati & Liem, 2021). But, on the other hand, up to 69% of consumers buy honey from a reliable farmer in an ordinary plastic container without any labellings (Roman et al., 2013). In this case, the producer is the most important criteria, and the package design is totally ignored. Nevertheless, some authors emphasize the role of the packaging material in the decision to buy honey. Many consumers prefer packaging made of smooth glass and react negatively to plastic packaging as the result of its unaesthetic appearance (Hazuchová et al., 2018). As it was underlined by Khaoula et al. (2019), glass packaging also signals about the high quality. In line with these observations, Nascimento et al. (2021) report that buyers believe that honey in a glass jar is more difficult to adulterate.

*Price.* Honey is considered as an expensive food product: there is a fraction of consumers that cannot afford buying honey frequently due to their financial situation. The cost of honey provokes doubts in the purchase decision-making process, and determines the search for a budget alternative or refusal to purchase this product by low-income population. However, previous research has demonstrated that there are certain types of honey for which consumers are willing to pay more than the average market price: organic and local origin honey (Vapa-Tankosic et al., 2020). The country of origin is an important aspect in the choice of honey because of the adulteration prevalence in the global market. In the study of Guziy et al. (2017), the respondents from Slovakia were worried about the quality of the imported honey and prefered to buy honey of local origin. In line with this finding, Zeng et al. (2023) report that almost 95% of chineese consumers prefer to buy local honey due to the fear of counterfeit foreign honey. The willingness to pay for local honey increases if consumers are provided with information about honey laundering in the imported honey (Ritten et al., 2019). In general, the willingness to pay for local or organic honey depends on consumers' socio-demographic characteristics. According to Vapa-Tankosic et al. (2020), women are more willing to pay for the organic honeythan men, and people with higher income levels value local products more. Willingness to pay for honey also increases with age and the level of education (Zeng et al., 2023).

Thus, in preference analysis, researchers often focus on the effect of externalized product characteristics (price, packaging material, point of sale, location of honey producer), while consumers decide to re-purchase honey after tasting the product and evaluating internalized characteristics (taste, aroma, texture, etc.). Here we aim to consider how the sensory characteristics of honey influence the perceived quality of the product.

## 2.2. Sensory characteristics of honey

There is a large variety of sensory characteristics that could be used to describe the flavor "portrait" of honey. Producers predominantly operationalize most of them for expert evaluation of the product quality. On the contrary, we were interested in product properties that can be identified directly by consumers: to better understand consumer behavior and to increase the willingness to pay, it is necessary to find out the terms that consumers may use to describe different honey samples. According to the existing literature, the frequently encountered nohey sensory characteristics include color and odor intensity, texture (liquid, crystalline), florality, fruitiness, degree of waxy, chemistry, fermentability, bitterness, astringency, sourness and acidity, mouthfeel, and the degree of aftertaste (Hunter et al., 2021).

*Texture.* When speaking about honey texture, we reffer to the perceived degree of crystallization, subdividing into liquid, creamy and crystallized consistencies. Traditionally, customers prefer honey with a liquid texture: it seems to be sweeter and more delicate. As honey crystallizes, the sweetness decreases, while the firmness and graininess increases (Piana et al., 2013). It is believed that consumers trust more liquid honey because they perceive it to be fresher (Šedík et al., 2023). On the other hand, in some studies, respondents prefer creamy to liquid honey (Khaoula et al., 2019).

*Aroma and Flavor.* Counterfeit honey usually has a less intense aroma, so consumers may focus on this indicator in a honey selection (Šedík et al., 2018). The sweetness can be identified as the most attractive characteristics, while most of the other sensory characteristics, according Hunter et al. (2021) have a negative impact on the attractiveness of honey. Furthermore, Hunter et al. (2021) establish a direct link between sweetness and the perceived amount of sugar in the product, suggesting that honey may be viewed by consumers as a hedonic product.

The exact set of sensory attributes used to describe honey perception may vary between regions and countries. For example, "jaggery-like" term could be used by Indian consumers to reffer to aroma associated with unrefined brown sugar made from palm sap (Anupama et al., 2003), while "animal-like" stands for a honeydew flavor in the Estonian study by Kivima et al. (2021). As a part of our behavioral experiment, we considered an additional characteristic that is familiar to Russian honey consumers, namely "peppercorn" (the tickling sensation in the throat after honey tasting).

## 3. Materials and method

### 3.1. Participants

A total of 25 healthy respondents (16 female / 9 male) have participated in the behavioral experiment. The respondents were recruited via snowball sampling through social media and selected according to the following criteria: 1) being from 25 to 80 years old; 2) without allergies to honey and other apiculture products; 3) absense of other diseases or neurological states that prevent honey degustation. Each participant had an experience of buying honey in a supermarket at least once. No restrictions regarding gender, education level or profession were applied. The majority of respondents represented young people aged 18-25 years (28%) and 26-35 years

(24%). People aged 46-65 years made up 40% of the sample in total. 68% of respondents had a higher education level, 56% of the sample were employees. All respondents provided their informed consent to participate in the study and received a gratification of 250 rubles after successful completion of the experiment.

**3.2. Experimental design and stimulus materials**

The behavioral experiment included three distinctive stages and lasted 1 hour 15 minutes in total.

*Preparatory stage.* Two series of behavioral experiments began with the instruction of the participants. Moderators explained the purpose and the course of the experiment, warned participants about potential allergic reactions to honey (including pollen of individual plants). At this stage, respondents also filled out paper-based surveys with questions about their socio-demographic characteristics, usual behavior regarding buying and consuming honey, the current level of satiety and their attitudes towards perceived health and hedonic food characteristics using Health and Taste attitide scale (Roininen & Tourila, 1999).

*Degustation.* At this stage respondents tasted 7 different samples of flower honey and evaluated each of them using pre-defined 21 sensory characteristics (the list of characteristics is given in the next subsection). Participants were free to define the order of samples and could correct their answers in the questionnaire form during the tasting. They were instructed to rinse their mouths with clean still water between tasting different samples. Honey samples weighted 50g each and were served in transparent plastic jars together with wooden tasting sticks. Respondents tasted as much honey as they considered necessary for grading and the time for tasting was not limited. During the tasting stage, respondents communicated only with moderators and did not share their opinion about the samples with other participants.

*Closed auction.* At the end of the behavioral experiment, participants received a gratification of 250 rubles and were offered to participate in a closed-type auction according to the Becker-DeGroot-Marschak method (Newton-Fenner et al., 2023). Each participant wrote on a piece of paper the amount he/she was willing to pay for a 100 g. glass jar of local natural honey (sample #7). In each of the two expirental sessions the respondent suggesting the maximum price received a jar of such honey, while his/her final remuneration was deducted accordingly.

**3.3. Questionnaire**
The survey filled out during the tasting stage contained the three following parts.
The first part included of a list of various sensory characteristics describing honey taste, aroma, and textural properties. Respondents evaluated the degree of manifestation of the characteristics in the tasting samples on a scale from 0 to 6, where 0 — complete inconsistency of the taste, aroma or texture characteristic of honey, 6 — maximum manifestation of the characteristics in the sample. For example, if respondents felt that the honey sample exhibited a very strong sweetness compared to other samples, they could give the sample a score of 6 out of 6 on the sweetness scale. A total of 21 sensory characteristics were considered:
1. color intensity
2. odor intensity

3. taste intensity
4. presence of foreign flavors
5. crystallization
6. florality
7. fruitiness
8. berryness
9. herbivory
10. woodiness
11. spiciness
12. tangibility of the taste or odor of wax in honey
13. the sensation of artificial sugar
14. sweetness
15. bitterness
16. sourness
17. peppercorn (the tickling sensation in the throat after the honey tasting)
18. duration of the aftertaste
19. "honey plant", the accordance between the expected sensation of honey and the actual taste of the main component
20. tartness
21. astringency

In the second part of the questionnaire respondents were asked to rate the likeability of the tasted honey samples on a scale from 0 to 6 and the perceived quality of honey on a scale from 1 to 7. Participants could also comment on their scores in a free text format if they were willing to explain very high or low scores.

The final part of the questionnaire assessed participants' willingness to pay for honey as a food product. Respondents assumed how much a 250-g. glass jar of a given honey sample may cost in a supermarket and indicated how much they were would be willing to pay for it.

### 3.3. Factor analysis

Grouping 21 sensory characteristics of honey into a smaller number of comprehensive features may be considered as a dimensionality reduction task. One of the classic approaches for this kind of tasks is an exploratory factor analysis (EFA). The term EFA is used to refer to two models that differ in purpose and computation: specifically, principal components analysis (PCA) and common factor analysis (Watkins, 2018). Here, we have applied a common factor analysis due to several reasons. First, it may better represent hidden interconnections in tha dataset compared to PCA (Gorsuch, 1990; Widaman, 1993). Secondly, PCA and common factor analysis yield very similar results if the identical number of factors (components) has been chosen (Velicer & Jackson, 1990). Thirdly, this type of analysis has already been successfully used in food preference studies (Sautron et. al, 2015). The common factor analysis and identification of the number of factors with "parallel" analysis (Horn, 1965) were performed with R programming language. After conducting common factor analysis the most relevant items from each factor were selected based on a criteria for loadings (equal to 0,4 or greater). The sustainability of the factors were checked by estimation of four linear regressions.

## 4. Results

### 4.1. Behavioral experiment

Despite the fact that 15 out of 25 respondents predominantly buy honey from familiar beekeepers, 44% of participants purchased honey in small volumes (less than 1 kg.), while volumes more than 3 kg. were purchased by 16% of respondents: all of them represent the age cohort about 56-65 years old. This phenomenon emphasizes the relevance of identifying the competitive advantages of the product for honey producers since small packages of honey can attract attention on the supermarket shelves, especially of the audience under 45 years.

Among significant positive correlations, we noted the relationship between frequency of honey consumption and purchase volume ($r_s = 0.41$), marital status and frequency of honey consumption ($r_s = 0.46$), having children and purchase volume ($r_s = 0.462$), having children and average willingness to pay across all honey samples ($r_s = 0.569$), frequency of honey consumption and average likeability across all the samples ($r_s = 0.51$). The degree of satiety at the beginning of the experiment had no significant effect on honey perception.

The closed auction stage ended successfully in both experimental sessions. In the first case, the maximal price offered for a 100g. jar of honey was equal to 250 rubles, which corresponded to the remuneration for participation in the experiment. In the second session, the highest offered price was 350 rubles. The distribution of the amounts offered by the respondents in the closed auction is presented in the Fig.1. According to the demand curve, the maximum revenue for the producer in this case may be reached at the price of 150 rubles per 100 g. of honey.

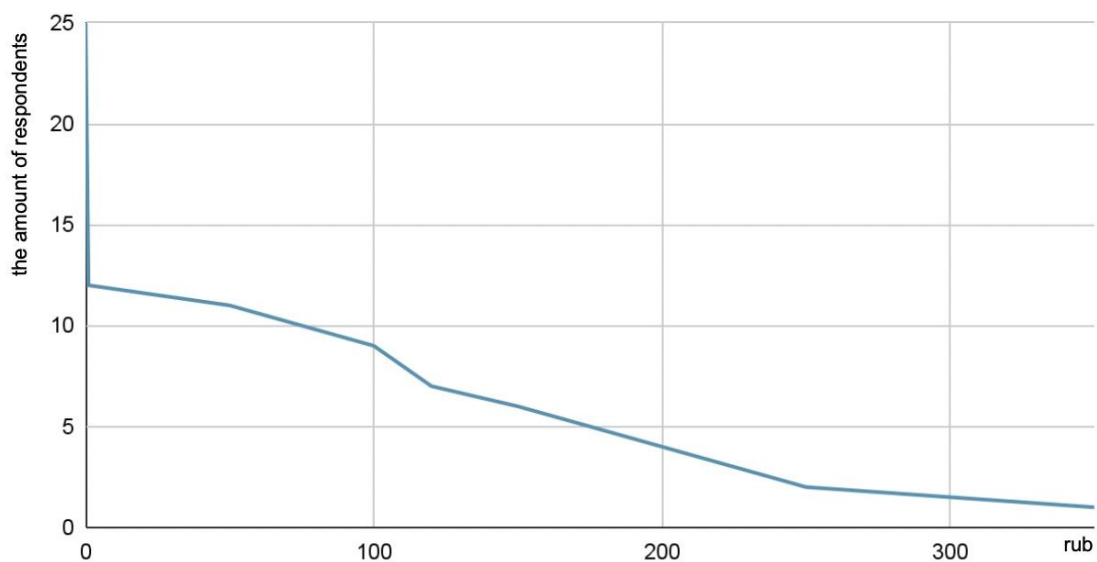

Figure 1. Demand curve for the closed auction

The experimental results demonstrated significant correlations between sensory characteristics of honey and the dependent variables: perceived quality and respondents' willingness to pay. The accordance between the expected sensation of honey and the actual taste of the main component, the intensity of flavor, the duration of aftertaste, and the tickling sensation in the throat after the honey tasting had a positive effect on the perceived quality of honey, while the sensation of extraneous flavors and added sugar shows the opposite trend. Interestingly, neither crystallization nor degree of sweetness influenced consumers' perception of honey samples ($|r_s|<0.06$).

The correlation matrix reveals the presence of relationships between different sensory characteristics: e.g. fruitiness and berryness ($r_s = 0.54$), woodiness and herbivory ($r_s = 0.44$), bitterness and "peppery" ($r_s = 0.59$). Thus, it seemed possible to combine the 21 sensory characteristics into several groups. Such grouping may reduce the number of components necesarry for describing the flavor "portrait" of honey and may help to make future honey perception studies less time-consuming and labor-intensive.

**4.2. Grouping sensory characteristics**

The results of the "parallel" analysis suggest that the number of final factors should be equal to eight (Fig.2). The resulting factors include (a name for each factor is suggested based on the highest factor loadings scores) :

1. Tartness
2. Absence of extraneous odors and components
3. Fruit and berry flavor
4. Peppercorn (the tickling sensation in the throat after the honey tasting)
5. Florality
6. Intensity of sweet flavor
7. Sourness
8. Sensation of taste, odor, or texture of the wax.

The loadings for each of these factors are given in Tab. 1. These factors were used as independent variables in regression analysis of the willingness-to-pay, likeability, percieved quality and retail price. The results of regression analysis are presented in Tab.2

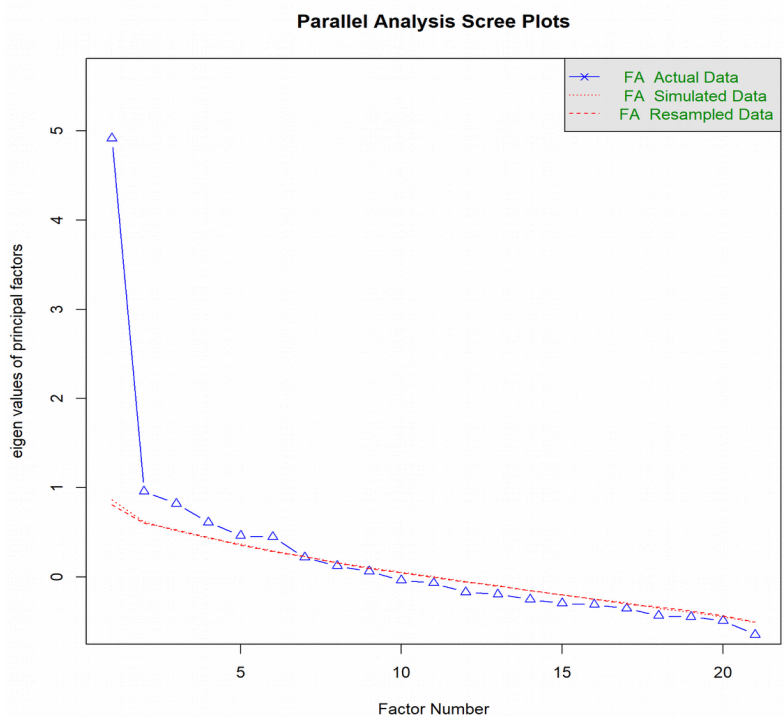

Figure 2. Representation of Horn's "parallel" analysis on the constructed dataset.

Table.1 The results of common factor analysis

| | Factor1 | Factor2 | Factor3 | Factor4 | Factor5 | Factor6 | Factor7 | Factor8 |
|---|---|---|---|---|---|---|---|---|
| Color intensity | | | | | | | | |
| Odor intensity | | | | | | | | |
| Taste intensity | | | | | | 0.567 | | |
| Foreign tastes | | -0.709 | | | | | | |
| Crystallization | | | | | | | | 0.427 |
| Florality | | | | | 0.932 | | | |
| Fruitiness | | | 0.963 | | | | | |
| Berry | | | 0.604 | | | | | |
| Herbivory | | | | | | | | |
| Woodiness | 0.438 | | | | | | | |
| Spice | 0.620 | | | | | | | |
| Wax | | | | | | | | 0.679 |
| Artificial sugar | | -0.567 | | | | | | |
| Sweetness | | | | | | 0.678 | | |
| Bitterness | 0.464 | | | 0.458 | | | | |
| Sourness | | | | | | | 0.438 | |
| Peppercorn | | | | 0.894 | | | | |
| Aftertaste | 0.653 | | | | | 0.466 | | |
| Honey plant | | 0.495 | | | | | | |
| Tartness | 0.764 | | | | | | | |
| Astringency | 0.530 | | | | | | | |

Table2. Estimated coefficients of four linear regressions

| | WTP | Quality | Likeability | Price |
|---|---|---|---|---|
| Factor1 | 36.872*** | 0.680*** | 0.567*** | 27.304*** |
| | (9.427) | (0.072) | (0.091) | (8.644) |
| Factor2 | 36.354*** | 0.923*** | 0.796*** | 6.503 |
| | (8.936) | (0.068) | (0.087) | (8.194) |
| Factor3 | 26.583** | 0.175** | 0.291*** | 29.610*** |
| | (10.568) | (0.081) | (0.102) | (9.690) |
| Factor4 | 17.184 | 0.259*** | 0.136 | 7.962 |
| | (10.470) | (0.080) | (0.101) | (9.600) |
| Factor5 | 7.471 | 0.321*** | 0.364*** | -1.376 |
| | (10.507) | (0.080) | (0.102) | (9.634) |
| Factor6 | 16.858* | 0.261*** | 0.198** | 17.434** |
| | (8.905) | (0.068) | (0.086) | (8.166) |
| Factor7 | 6.180 | 0.220*** | 0.223*** | 9.982 |
| | (8.681) | (0.066) | (0.084) | (7.960) |
| Factor8 | 15.451* | 0.007 | 0.022 | 2.414 |
| | (8.790) | (0.067) | (0.085) | (8.060) |
| Constant | 205.031*** | 4.484*** | 3.629*** | 278.805*** |
| | (10.613) | (0.081) | (0.103) | (9.731) |
| Observations | 159 | 159 | 159 | 159 |
| $R^2$ | 0.216 | 0.654 | 0.480 | 0.140 |
| Adjusted $R^2$ | 0.174 | 0.636 | 0.452 | 0.094 |
| Residual Std. Error (df = 150) | 133.821 | 1.020 | 1.297 | 122.704 |
| F Statistic (df = 8; 150) | 5.165*** | 35.487*** | 17.282*** | 3.044*** |
| Note: | | | | *p<0.1; **p<0.05; ***p<0.01 |

## 5. Discussion

The first step in assessing the perceived quality of honey was to conduct a correlation analysis of the survey data from the behavioral experiment on the honey sensory characteristics. Thus, the correlation analysis demonstrated that the strongest influence on the perceived quality of honey is exerted by the accordance between the expected sensation of honey and the actual taste of the main component (with the corresponding value of Spearman correlation coefficient $r_s = 0.66$), duration of aftertaste ($r_s = 0.56$) and tartness ($r_s = 0.52$). We have found the strongest negative relationship between the perceived quality of honey and the characteristics of artificial sugar sensitivity ($r_s = -0.57$) and foreign flavors ($r_s = -0.49$). It is noteworthy that crystallization, which according to some studies is often associated by consumers with added sugars (e.g., glucose-fructose syrup or maltose syrup), was not significantly correlated with either the artificial sugar palatability characteristic or perceived quality. Based on the results of the correlation analysis, we concluded that the most significant sensory characteristics of honey for Russian consumers are factors related to taste, which is in agreement with general results of previous studies (Hunter et al., 2021).

Further, taking into account the complexity of the task of identifying individual flavor characteristics by ordinary consumers, we conducted an explanatory factor analysis, so that 21 sensory characteristics of honey were combined into 8 factors. Given the high proportion of the explained variance, it can be concluded that these eight the more extensive descriptive characteristics are sufficient to form a complete flavor portrait of honey. First of all, reducing the number of analyzed characteristics allows to simplify the tasting procedure for respondents who were not professional tasters capable of identifying the subtlest differences in numerous characteristics. Secondly, this may help to reduce of the experiment duration, because the respondents will have to evaluate only 8 characteristics instead of 21.

**Acknowledgements:**
The article was prepared within the framework of the Basic Research Program at HSE University. The authors thank Anton Astafiev from the Apiary of the Astafiev family "Lyubomedovo" for providing honey samples.